\documentclass[conference]{IEEEtran}
\usepackage{amssymb}

\makeatletter
\def\ps@headings{%
\def\@oddhead{\mbox{}\scriptsize\rightmark \hfil \thepage}%
\def\@evenhead{\scriptsize\thepage \hfil \leftmark\mbox{}}%
\def\@oddfoot{}%
\def\@evenfoot{}}
\makeatother
\usepackage{amsfonts}
\usepackage{amssymb,color}
\usepackage{amsmath,epsfig}
\usepackage{rawfonts,graphicx,amsfonts,amssymb}
\usepackage{array}
\usepackage{mathrsfs}
\usepackage{cite}
\usepackage[normal]{subfigure}
\usepackage[ruled,vlined,linesnumbered]{algorithm2e}
\usepackage{footnote}
\usepackage{url}
\usepackage{multirow}
\usepackage{threeparttable}

\ifodd 0
\newcommand{\com}[1]{\textbf{\color{blue} (COMMENT: #1)}} 
\else

\newcommand{\com}[1]{}
\fi

\newtheorem{theorem}{Theorem}

\newtheorem{definition}{Definition}

\IEEEoverridecommandlockouts

\ifCLASSINFOpdf \else \fi

\begin{document}

\title{Effective Carrier Sensing in CSMA Networks under
Cumulative Interference}

%

\author{\IEEEauthorblockN{Liqun Fu, Soung Chang Liew, Jianwei
Huang}
\IEEEauthorblockA{Department of Information Engineering\\
The Chinese University of Hong Kong\\
Shatin, New Territories, Hong Kong\\
Email: \{lqfu6,soung,jwhuang\}@ie.cuhk.edu.hk}

\thanks{This work was supported by two
Competitive Earmarked Research Grants (Project Number 414507 and
Project Number 412308) established under the University Grant
Committee of the Hong Kong Special Administrative Region, China, the
Direct Grant (Project Number C001-2050398) of The Chinese University
of Hong Kong, and the National Key Technology R\&D Program (Project
Number 2007BAH17B04) established by the Ministry of Science and
Technology of the People's Republic of China. }

}


\maketitle

\begin{abstract}

This paper proposes and investigates the concept of a \emph{safe
carrier-sensing range} that can guarantee interference-safe (also
termed hidden-node-free) transmissions in CSMA networks under the
cumulative interference model. Compared with the safe
carrier-sensing range under the commonly assumed but less realistic
pairwise interference model, we show that the safe carrier-sensing
range required under the cumulative interference model is larger by
a constant multiplicative factor. For example, if the SINR
requirement is $10dB$ and the path-loss exponent is $4$, the factor
is $1.4$. The concept of a safe carrier-sensing range, although
amenable to elegant analytical results, is inherently not compatible
with the conventional power-threshold carrier-sensing mechanism
(e.g., that used in IEEE 802.11). Specif\/ically, the absolute power
sensed by a node in the conventional mechanism does not contain
enough information for it to derive its distances from other
concurrent transmitter nodes. We show that, fortunately, a
carrier-sensing mechanism called Incremental-Power Carrier-Sensing
(IPCS) can realize the carrier-sensing range concept in a simple
way. Instead of monitoring the absolute detected power, the IPCS
mechanism monitors every increment in the detected power. This means
that IPCS can separate the detected power of every concurrent
transmitter, and map the power prof\/ile to the required distance
information. Our extensive simulation results indicate that IPCS can
boost spatial reuse and network throughput by more than $60\%$
relative to the conventional carrier-sensing mechanism. Last but not
least, IPCS not only allows us to implement our safe carrier-sensing
range, it also ties up a loose end in many other prior theoretical
works that implicitly assume the use of a carrier-sensing range
(safe or otherwise) without an explicit design to realize it.

\end{abstract}

\begin{keywords}
carrier-sensing range, cumulative interference model, CSMA, WiFi,
IEEE 802.11, SINR constraints, spatial reuse.
\end{keywords}

\IEEEpeerreviewmaketitle

\section{Introduction and Overview}

Due to the broadcast nature of wireless channels, signals
transmitted over wireless links can mutually interfere with each
other. How to optimize spatial reuse and network throughput under
such mutual interferences has been an intensely studied issue in
wireless networking. In particular, it is desirable to allow as many
links as possible to transmit together in an interference-safe (or
collision-free) manner. The problem of interference-safe
transmissions under the coordination of a centralized TDMA
(Time-Division Multiple-Access) scheduler has been well studied
(e.g., see
\cite{Gurashish,ephremides2,gsharma,oswald,thomasro,Padhye}). Less
well understood is the issue of interference-safe transmissions
under the coordination of a distributed scheduling protocol.

The CSMA (Carrier-Sense Multiple-Access) protocol, such as IEEE
802.11, is the most widely adopted distributed scheduling protocol
in practice. As the growth of 802.11 network deployments continues
unabated, we are witnessing an increasing level of mutual
interference among transmissions in such networks. It is critical to
establish a rigorous conceptual framework upon which effective
solutions to interference-safe transmissions can be constructed.

Within this context, this paper has three major contributions listed
as follows (more detailed overview is given in the succeeding
paragraphs):
\begin{enumerate}
\item{We propose the concept of a \emph{safe carrier-sensing range} that
can guarantee interference-safe transmissions in CSMA networks under
the \emph{cumulative interference model}.}
\item{We show that the concept is implementable using a very simple
Incremental-Power Carrier-Sensing (IPCS) mechanism.}
\item{We demonstrate that implementation of safe carrier-sensing range
under IPCS can signif\/icantly improve spatial reuse and network
throughput as compared to the conventional absolute-power carrier
sensing mechanism.}
\end{enumerate}

Regarding 1), this paper considers the cumulative interference model
(also termed physical interference model in \cite{prgupta}), in
which the interference at a receiver node $i$ consists of the
cumulative power received from all the other nodes that are
currently transmitting (except its own transmitter). This model is
known to be more practical and much more diff\/icult to analyze than
the widely studied pairwise interference model (also termed the
protocol interference model in \cite{prgupta}) in the literature.
Under the cumulative interference model, a set of simultaneously
transmitting links are said to be interference-safe if the SINRs
(Signal-to-Interference-plus-Noise Ratios) at the receivers of all
these links are above a threshold. Given a set of links
$\mathcal{L}$ in the network, there are many subsets of links,
$\mathcal{S}\subset \mathcal{L}$, that are interference-safe. The
set of all such subsets $\mathcal{F}=\{\mathcal{S}\mid \text{the
SINR requirements of all links are satisf\/ied}\}$ constitutes the
feasible interference-safe state space. For centralized TDMA, all
subsets are available for scheduling, and a TDMA schedule is
basically a sequence $(\mathcal{S}_t)_{t=1}^n$ where each
$\mathcal{S}_t\in\mathcal{F}$. For CSMA, because of the random and
distributed nature of the carrier-sensing operations by individual
nodes, the simultaneously transmitting links $\mathcal{S}^{CS}$ may
or may not belong to $\mathcal{F}$. Let
$\mathcal{F}^{CS}=\{\mathcal{S}^{CS}\mid \text{simultaneous
}\text{transmissions of links in } \mathcal{S}^{CS} \text{ are
allowed by }\\\text{the carrier-sensing operation}\}$. The CSMA
network is said to be interference-safe if
$\mathcal{F}^{CS}\subseteq \mathcal{F}$. This is also the condition
for the so-called hidden-node free operation \cite{LiBin}. However,
this issue was studied under the context of an idealized pairwise
interference model \cite{LiBin} rather than the practical cumulative
interference model of interest here. In this paper, we show that if
the carrier-sensing mechanism can guarantee that the distance
between every pair of transmitters is separated by a \emph{safe
carrier-sensing range}, then $\mathcal{F}^{CS}\subseteq \mathcal{F}$
can be guaranteed and the CSMA network is interference-safe even
under a cumulative interference model. We believe that the safe
carrier-sensing range established in this paper is a tight
upperbound and achieves good spatial reuse. Another issue is how to
implement the concept of safe carrier-sensing range in practice.


This brings us to 2) above. In traditional carrier sensing based on
power threshold (e.g., that of the \emph{basic mode} in IEEE
802.11), the absolute power received is being monitored. This power
consists of the sum total of powers received from all the other
transmitters. It is impossible to infer from this absolute power the
exact separation of the node from each of the other transmitters.
This leads to subpar spatial reuse. Fortunately, we show that a
simple mechanism that monitors the incremental power changes over
time, IPCS, will enable us to map the power prof\/ile to the
required distance information. We believe that this contribution,
although simple, is signif\/icant in that it shows that the
theoretical concept of \emph{safe carrier-sensing range} can be
implemented rather easily in practice. It also ties up a loose end
in many other prior theoretical works that implicitly assume the use
of a carrier-sensing range (safe or otherwise) without an explicit
design to realize it. That is, IPCS can be used to implement the
required carrier-sensing range in these works, not just our
\emph{safe carrier-sensing range} here. Without IPCS, and using only
the conventional carrier-sensing mechanism, the results in these
prior works would have been overly optimistic. Given the
implementability of safe carrier-sensing range, the next issue is
how tight the simultaneously transmitting nodes can be packed.


This brings us to 3) above. In the conventional carrier sensing
mechanism, in order that the detected absolute power is below the
carrier-sensing power threshold, the separation between a newly
active transmitter and other existing active transmitters must
increase progressively as the number of concurrent transmissions
increases. That is, the cost of ensuring interference-safe
transmissions becomes progressively higher and higher in the
``packing process''. This reduces spatial reuse and the overall
network throughput. Fortunately, with IPCS, the required separation
between any pair of active transmitters remains constant as the
\emph{safe carrier-sensing range} which is independent of the number
of concurrent transmissions. Indeed, our simulation results indicate
that compared to the conventional carrier-sensing mechanism, IPCS
mechanism improves the spatial reuse and the network throughput by
more than $60\%$.




\subsection{Related Work}\label{relatedwork}

In the literature, most studies on carrier sensing (e.g.,
\cite{KXu,LiBin,PCNg,libinhdfvcs,SXu,Vasan}) are based on the
pairwise interference model. For a link under the pairwise
interference model, the interferences from the other links are
considered one by one. If the interference from each of the other
links on the link concerned does not cause a collision, then it is
assumed that there is no collision overall. Ref. \cite{LiBin}
established the carrier-sensing range required to prevent
hidden-node collisions in CSMA networks under the pairwise
interference model. The resulting carrier-sensing range is too
optimistic and can not eliminate hidden-node collisions if the more
accurate cumulative interference model is adopted instead.

A number of recent papers studied the CSMA networks under the
cumulative interference model (e.g.,
\cite{chichau,TaeSuk,TingYu,Liqun}). An earlier unpublished
technical report of ours \cite{Liqun} derived the safe
carrier-sensing range under the cumulative interference model. The
technical report, however, did not include the IPCS realization
presented in this paper. Neither did Ref.
\cite{chichau,TaeSuk,TingYu} address the implementation of a
carrier-sensing range based on power detection. Ref. \cite{chichau}
studied the asymptotic capacity of large-scale CSMA networks with
hidden-node-free designs. The focus of \cite{chichau} is on
``order'' result rather than ``tight'' result. For example, if
$\gamma_0=10dB$ and $\alpha=4$, the safe carrier-sensing range
derived in \cite{chichau} is $8.75d_{\max}$. In this paper, we show
that setting the safe carrier-sensing range to $5.27d_{\max}$ is
enough to prevent hidden-node collisions.

The authors in \cite{TaeSuk,TingYu} attempted to improve spatial
reuse and capacity by tuning the transmit power and the
carrier-sensing range. Although the cumulative interference model is
considered in \cite{TaeSuk,TingYu}, spatial reuse and capacity are
analyzed based on carrier-sensing range. In particular, they assumed
that the transmitters of concurrent transmission links can be
uniformly packed in the network. As discussed in this paper, such
uniform packing can not be realized using the current 802.11
carrier-sensing mechanism. Therefore, the results in
\cite{TaeSuk,TingYu} are overly optimistic without an appropriate
carrier-sensing mechanism. IPCS f\/ills this gap so that the
theoretical results of \cite{TaeSuk,TingYu} remain valid. We
summarize the key related models and results in the literature in
Table \ref{Relatework}\footnote{This paper focuses on the
incremental-power carrier-sensing (IPCS) mechanism under the
cumulative interference model. But IPCS proposed in this paper can
also deal with the pairwise interference model.}.

\begin{table}[tb]
\caption{Summary of the Related Work}
\begin{center}
\renewcommand{\arraystretch}{1.3}
\footnotesize
\begin{tabular}{m{65pt}||m{68pt}|m{68pt}}
\hline%
\centering Interference Models & \centering Pairwise Interference
Model &\centering Cumulative Interference Model
\tabularnewline%
\hline\hline%
\centering Absolute power carrier sensing  & \centering many (e.g.,
\cite{LiBin,KXu})  &\centering \cite{TaeSuk,TingYu}
\tabularnewline%
\hline \centering Incremental power carrier sensing  &\centering
\textbf{This paper}&\centering \textbf{This paper}
\tabularnewline%
\hline
\end{tabular}
\end{center}
\label{Relatework}
\end{table}

The rest of this paper is organized as follows. Section \ref{system}
presents the cumulative interference model and the carrier sensing
mechanism in the current 802.11 protocol. Section \ref{safecsphy}
derives the safe carrier-sensing range that successfully prevents
the hidden-node collisions under the cumulative interference model.
Section \ref{IPCS} presents the IPCS mechanism. Section
\ref{simulation} evaluates the performance of IPCS in terms of
spatial reuse and network throughput. Section \ref{conclusion}
concludes this paper.

\section{System Model}\label{system}

\subsection{Cumulative Interference Model}\label{Model}

We represent links in a wireless network by a set of distinct and
directed transmitter-receiver pairs $\mathcal{L}=\{l_i, 1\le i \le
{\left| \mathcal{L} \right|}\}$. Let $\mathcal{T}=\{T_i, 1\le i \le
{\left| \mathcal{L} \right|}\}$ and $\mathcal{R}=\{R_i, 1\le i \le
{\left| \mathcal{L} \right|}\}$ denote the set of transmitter nodes
and the set of receiver nodes, respectively. A receiver decodes its
signal successfully if and only if the received
Signal-to-Interference-plus-Noise Ratio (SINR) is above a certain
threshold. We adopt the cumulative interference model, where the
interference is the sum of the received powers from all transmitters
except its own transmitter. We assume that radio signal propagation
follows the log-distance path model with path loss exponent
$\alpha>2$. The path gain $G(T_i,R_j)$ from transmitter $T_i$ to
receiver $R_j$ follows a geometric model:
\begin{equation}
G(T_i,R_j)={d(T_i,R_j)}^{-\alpha}, \label{channelm}\nonumber
\end{equation}
where $d(T_i,R_j)$ is the Euclidean distance between nodes $T_i$ and
$R_j$.

In 802.11, each packet transmission on a link $l_i$ consists of a
DATA frame in the forward direction (from $T_i$ to $R_i$) followed
by an ACK frame in the reverse direction (from $R_i$ to $T_i$). The
packet transmission is said to be successful if and only if both the
DATA frame and the ACK frame are received correctly. Let
$\mathcal{L}'$ ($\mathcal{L}''$) denote the set of links that
transmit concurrently with the DATA (ACK) frame on link $l_i$. Under
the cumulative interference model, a successful transmission on link
$l_i$ needs to satisfy the following conditions:
\begin{equation} \frac{P_t\cdot
G(T_i ,R_i )}{N + \sum\limits_{l_j \in \mathcal{L}'}{P_t\cdot G(S_j
,R_i )} } \ge {\gamma}_0, \quad \text{(DATA frame)}\label{SINR1}
\end{equation}
and
\begin{equation} \frac{P_t\cdot
G(R_i ,T_i )}{N + \sum\limits_{l_j \in \mathcal{L}''}{P_t\cdot G(S_j
,T_i )} } \ge {\gamma}_0, \quad \text{(ACK frame)}\label{SINR2}
\end{equation}
where $P_t$ is the transmit power, $N$ is the average noise power,
and $\gamma_0$ is the SINR threshold for correct reception. We
assume that all nodes in the network use the same transmit power
$P_t$ and adopt the same SINR threshold $\gamma_0$. For a link $l_j$
in $\mathcal{L}'$ or $\mathcal{L}''$, $S_j$ represents the sender of
link $l_j$, which can be either $T_j$ or $R_j$. This is because
either DATA or ACK transmission on link $l_j$ will cause
interference to link $l_i$.


\subsection{Existing Carrier Sensing Mechanism in 802.11}\label{carriersense}

If there exists a link $l_i\in\mathcal{L}$ such that not both
\eqref{SINR1} and \eqref{SINR2} are satisf\/ied, this means there is
collision in the network. In 802.11, carrier sensing is designed to
prevent collision due to simultaneous transmissions that cause the
violation of either \eqref{SINR1} or \eqref{SINR2}. In this paper,
we assume carrier sensing by energy detection. Consider a link
$l_i$. If transmitter $T_i$ senses a power $P^{CS}(T_i)$ that
exceeds a power threshold $P_{th}$, i.e.,
\begin{equation}
P^{CS}(T_i)>P_{th}, \label{eqPcs}
\end{equation}
then $T_i$ will not transmit and its backoff countdown process will
be frozen. This will prevent the DATA frame transmission on $l_i$.

In most studies of 802.11 networks, the concept of a carrier-sensing
range $CSR$ is introduced. The carrier-sensing range $CSR$ is mapped
from the carrier-sensing power threshold $P_{th}$:
\begin{equation}
CSR=\left(\frac{P_t}{P_{th}}\right)^{\frac{1}{\alpha}}.\nonumber
\label{eqTth}
\end{equation}

Consider two links, $l_i$ and $l_j$. If the distance between
transmitters $T_i$ and $T_j$ is no less than the carrier-sensing
range, i.e.,
\begin{equation}
d(T_i,T_j)\geq CSR, \label{eqCSRange}
\end{equation}
then $T_i$ and $T_j$ can not carrier sense each other, and thus can
initiate concurrent transmissions between them. The pairwise
relationship can be generalized to a set of links
$\mathcal{S}^{CS}\subseteq \mathcal{L}$. If the condition in
\eqref{eqCSRange} is satisf\/ied by all pairs of transmitters in set
$\mathcal{S}^{CS}$, then all links in $\mathcal{S}^{CS}$ can
transmit concurrently.

Setting an appropriate carrier-sensing range is crucial to the
performance of 802.11 networks. If $CSR$ is too large, spatial reuse
will be unnecessarily limited. If $CSR$ is not large enough, then
hidden-node collisions may occur. The underlying cause of
hidden-node collisions are as follows. A number of transmitters
transmit simultaneously because condition \eqref{eqCSRange} is
satisf\/ied by all pairs of the transmitters. However, there is at
least one of the links does not satisfy either \eqref{SINR1} or
\eqref{SINR2}. As a result, collisions happen and the carrier
sensing mechanism is said to have failed in preventing such
collisions.

We now def\/ine a \emph{safe carrier-sensing range} that always
prevents the hidden-node collisions in 802.11 networks under the
cumulative interference model.
\begin{definition}[Safe-$CSR_{\text{cumulative}}$]
Let $\mathcal{S}^{ CS}\subseteq \mathcal{L}$ denote a subset of
links that are allowed to transmit concurrently under a
carrier-sensing range $CSR$. Let
$\mathcal{F}^{CS}=\{\mathcal{S}^{CS}\}$ denote all such subsets of
links in the network. A $CSR$ is said to be a
\emph{Safe-}$CSR_{\text{\emph{cumulative}}}$ if for any
$\mathcal{S}^{CS}\in \mathcal{F}^{CS}$ and for any link
$l_i\in\mathcal{S}^{CS}$, both conditions \eqref{SINR1} and
\eqref{SINR2}, with
$\mathcal{L}'=\mathcal{L}''=\mathcal{S}^{CS}\setminus \{l_i\}$, are
satisf\/ied.
\end{definition}


For analysis simplicity, we assume that the background noise power
$N$ is small compared with interference and thus can be ignored. We
will consider Signal-to-Interference Ratio (SIR) instead of SINR.

\section{Safe Carrier-sensing Range under Cumulative Interference
Model}\label{safecsphy}

In this section, we derive a suff\/icient threshold for
\emph{Safe-}$CSR_{\text{\emph{cumulative}}}$. When discussing the
hidden-node free design \cite{LiBin}, it is required that the
receivers are operated with the ``RS (Re-Start) mode'' (see Appendix
\ref{apprestartmode} for details). In the following discussion, we
also make the same assumption.

\begin{figure}[http]
\begin{center}
\includegraphics [height=1.8cm]{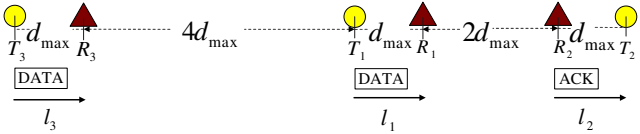}
\end{center}
\begin{center}
\vspace*{-0.20cm} \caption{Setting the carrier-sensing range as
\emph{Safe-}$CSR_{\text{\emph{pairwise}}}$ is insuff\/icient to
prevent hidden-node collisions under the cumulative interference
model} \label{pairinsuf}
\end{center}
\end{figure}

Ref. \cite{LiBin} studied the safe carrier-sensing range under the
\emph{pairwise interference model}. The threshold is given as
follows:
\begin{equation}
\emph{Safe-}CSR_{\text{\emph{pairwise}}}=\left({\gamma_0}^{\frac{1}{\alpha
}}+2\right)d_{\max}, \label{eqPthpair}
\end{equation}
where $d_{\max}=\mathop {\max }\limits_{l_i \in
\mathcal{L}}d(T_i,R_i)$ is the maximum link length in the network.
However, the pairwise interference model does not take into account
the cumulative nature of interferences from other links. The
threshold given in \eqref{eqPthpair} is overly optimistic and not
large enough to prevent hidden-node collisions under the
\emph{cumulative interference model}, as illustrated by the
three-link example in Fig. \ref{pairinsuf}.

In Fig. \ref{pairinsuf}, suppose that the SIR requirement
$\gamma_0=8$ and the path-loss exponent $\alpha=3$. According to
\eqref{eqPthpair}, it is enough to set the carrier-sensing range as
$\left({\gamma_0}^{\frac{1}{\alpha }}+2 \right)d_{\max}=4d_{\max}$
and the carrier sensing power threshold
$P_{th}=P_t\left(4d_{\max}\right)^{-3}= 0.0156 P_td_{\max}^{-3}$. In
Fig. \ref{pairinsuf}, there are three links: $l_1$, $l_2$, and $l_3$
with the same link length $d_{\max}$. The distance $d(R_1,R_2)$
equals $2d_{\max}$ and the distance $d(T_1,R_3)$ equals $4d_{\max}$.
Since the distance
$d(T_1,T_2)=4d_{\max}=\left({\gamma_0}^{\frac{1}{\alpha }}+2
\right)d_{\max}$, from \eqref{eqTth}, we f\/ind that $T_1$ and $T_2$
can simultaneously initiate transmissions since they can not carrier
sense each other. We can verify that the SIR requirements of both
DATA and ACK transmissions on $l_1$ and $l_2$ are satisf\/ied. This
means $l_1$ and $l_2$ can indeed successfully transmit
simultaneously.

Suppose that $l_3$ wants to initiate a transmission when $T_1$ is
sending a DATA frame to $R_1$ and $R_2$ is sending an ACK frame to
$T_2$. Transmitter $T_3$ senses a power $P^{CS}(T_3)$ given by
\begin{align}
P^{CS}(T_3)&=P_t\cdot(5d_{\max})^{-3}+P_t\cdot(8d_{\max})^{-3}\nonumber\\
&=0.00995\cdot P_td_{\max}^{-3}<P_{th}.\nonumber
\end{align}
This means that $T_3$ can not sense the transmissions on $l_1$ and
$l_2$, and can initiate a DATA transmission. However, when all these
three links are active simultaneously, the SIR at $R_1$ is
\begin{equation}
\frac{P_t(d_{\max})^{-3}}{{P_t(6d_{\max})^{-3}}+{P_t(2d_{\max})^{-3}}
}=7.714< {\gamma}_0.\nonumber
\end{equation}
This means the cumulative interference powers from $l_2$ and $l_3$
will corrupt the DATA transmission on $l_1$ due to the
insuff\/icient SIR at $R_1$. This example shows that setting the
carrier-sensing range as in \eqref{eqPthpair} is not suff\/icient to
prevent collisions under the cumulative interference model.



We next establish a threshold for
\emph{Safe-}$CSR_{\text{\emph{cumulative}}}$ so that the system will
remain safe under cumulative interference.




\begin{theorem}
\label{SafeCSRange} The setting
\begin{equation}
\emph{Safe-}CSR_{\text{\emph{cumulative}}}=(K+2)d_{\max}
\label{SafeCSRangeReq},
\end{equation}
where
\begin{equation}
K = \left( {6\gamma_0 \left( {1 + \left( {\frac{2}{\sqrt 3 }}
\right)^\alpha \frac{1}{\alpha - 2}} \right)}
\right)^{\frac{1}{\alpha }}. \label{KReq}
\end{equation}
is suff\/icient to ensure interference-safe transmissions under the
cumulative interference model.
\end{theorem}

\begin{proof}
The proof is given in Appendix \ref{theorem1proof}.
\end{proof}

Condition \eqref{SafeCSRangeReq} provides a suff\/iciently large
carrier-sensing range that prevents the hidden-node collisions in
CSMA networks. Therefore, there is no need to set a $CSR$ larger
than the value given in \eqref{SafeCSRangeReq}.

%


Let us compare \emph{Safe-}$CSR_\text{\emph{cumulative}}$ with
\emph{Safe-}$CSR_\text{\emph{pairwise}}$ with different values of
$\gamma_0$ and $\alpha$. For example, if $\gamma_0=10$ and
$\alpha=4$, which are typical for wireless communications,
\begin{align}
&\text{\emph{Safe-}}CSR_\text{\emph{pairwise}}=3.78\cdot d_{\max}
,\nonumber\\
&\text{\emph{Safe-}}CSR_\text{\emph{cumulative}}=5.27\cdot d_{\max}
\nonumber.
\end{align}
Compared with \emph{Safe-}$CSR_\text{\emph{pairwise}}$,
\emph{Safe-}$CSR_\text{\emph{cumulative}}$ needs to be increased by
a factor of $1.4$ to ensure successful transmissions under the
cumulative interference model.

Given a f\/ixed path-loss exponent $\alpha$, both
\emph{Safe-}$CSR_\text{\emph{pairwise}}$ and
\emph{Safe-}$CSR_\text{\emph{cumulative}}$ increase in the SIR
requirement $\gamma_0$. This is because the separation among links
must be enlarged to meet a larger SIR target. For example, if
$\alpha=4$, we have
\begin{align}
&\text{\emph{Safe-}}CSR_\text{\emph{pairwise}}=\left(2+\gamma_0^{\frac{1}{4}}\right)d_{\max}
,\nonumber\\
&\text{\emph{Safe-}}CSR_\text{\emph{cumulative}}=\left(2+\left(\frac{34}{3}\gamma_0\right)^{\frac{1}{4}}\right)d_{\max}
.\nonumber
\end{align}
The ratio of \emph{Safe-}$CSR_\text{\emph{cumulative}}$ to
\emph{Safe-}$CSR_\text{\emph{pairwise}}$ is
\begin{equation}
\mathop
\frac{\text{\emph{Safe-}}CSR_\text{\emph{cumulative}}}{\text{\emph{Safe-}}CSR_\text{\emph{pairwise}}}
=  \frac{2 + \left( {\frac{34}{3}\gamma_0 } \right)^{\frac{1}{4}}}{2
+ \gamma_0 ^{\frac{1}{4}}},\nonumber
\end{equation}
which is an increasing function of $\gamma_0$, and converges to a
constant as $\gamma_0$ goes to inf\/inity:
\begin{align}
\mathop {\lim }\limits_{\gamma_0 \to \infty }
\frac{\text{\emph{Safe-}}CSR_\text{\emph{cumulative}}}{\text{\emph{Safe-}}CSR_\text{\emph{pairwise}}}
&= \mathop {\lim }\limits_{\gamma_0 \to \infty } \frac{2 + \left(
{\frac{34}{3}\gamma_0 } \right)^{\frac{1}{4}}}{2 + \gamma_0
^{\frac{1}{4}}}\nonumber \\&= \left( {\frac{34}{3}}
\right)^{\frac{1}{4}} \approx 1.8348. \nonumber
\end{align}

\begin{figure}[t]
\begin{center}
\includegraphics [height=7.0cm]{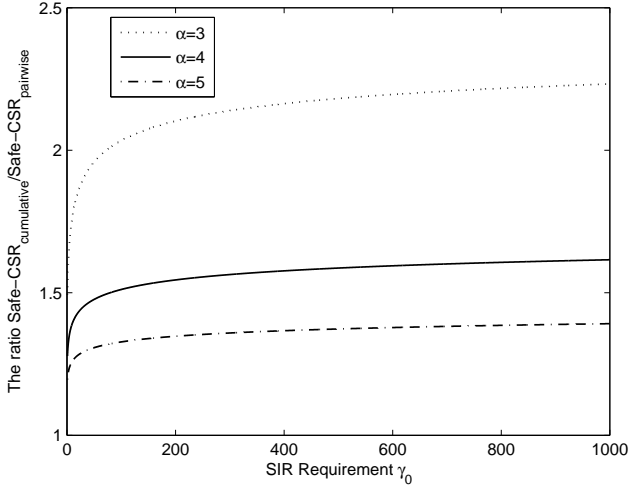}
\end{center}
\begin{center}
\vspace*{-0.20cm} \caption{The ratio of
\emph{Safe-}$CSR_\text{\emph{cumulative}}$ to
\emph{Safe-}$CSR_\text{\emph{pairwise}}$} \label{CSRangeratio}
\end{center}
\end{figure}

Fig. \ref{CSRangeratio} shows the ratio
$\frac{\text{\emph{Safe-}}CSR_\text{\emph{cumulative}}}{\text{\emph{Safe-}}CSR_\text{\emph{pairwise}}}$
as a function of the SIR requirements $\gamma_0$. Different curves
represent different choices of the path-loss exponent $\alpha$. The
ratio
$\frac{\text{\emph{Safe-}}CSR_\text{\emph{cumulative}}}{\text{\emph{Safe-}}CSR_\text{\emph{pairwise}}}$
increases when $\gamma_0$ increases or $\alpha$ decreases. For each
choice of $\alpha$, the ratio converges to a constant as $\gamma_0$
goes to inf\/inity. This shows that, compared with the pairwise
interference model, the safe carrier-sensing range under the
cumulative interference model will not increase arbitrarily.

\section{A Novel Carrier Sensing Mechanism}\label{IPCS}

We now discuss the implementation of
\emph{Safe-}$CSR_{\text{\emph{cumulative}}}$. We f\/irst describe
the diff\/iculty of implementing the safe carrier-sensing range in
\eqref{SafeCSRangeReq} using the existing physical carrier-sensing
mechanism in the current 802.11 protocol. Then, we propose a new
Incremental-Power Carrier-Sensing (IPCS) mechanism to resolve this
implementation issue.


\subsection{Limitation of Conventional Carrier-Sensing
Mechanism}\label{limitation}

\begin{figure}[http]
\begin{center}
\vspace*{0.20cm}
\includegraphics [height=6.4cm]{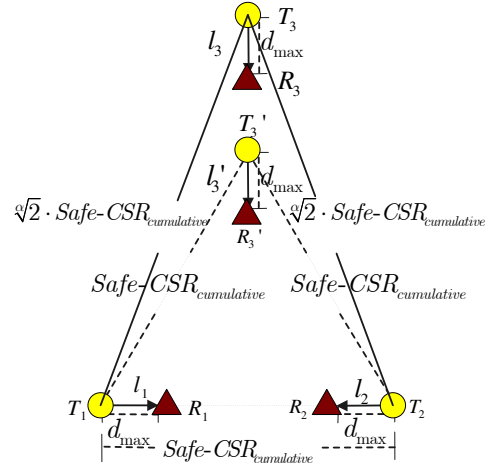}
\end{center}
\begin{center}
\vspace*{-0.20cm} \caption{Conventional carrier-sensing mechanism
will reduce the spatial reuse in 802.11 networks. Link $l_3$ is
placed based on the absolute power sensing mechanism in current
802.11, and link $l'_3$ is placed based on the
\emph{Safe-}$CSR_{\text{\emph{cumulative}}}$ as enabled by our IPCS
mechanism. } \label{limitspr} \vspace*{0.30cm}
\end{center}
\end{figure}

In the current 802.11 MAC protocol, given the safe carrier-sensing
range \emph{Safe-}$CSR_{\text{\emph{cumulative}}}$, the
carrier-sensing power threshold $P_{th}$ is set as
\begin{equation}
P_{th}=
P_t\cdot\left(\emph{Safe-}CSR_{\text{\emph{cumulative}}}
\right)^{-\alpha}.\label{abspower}
\end{equation}
Before transmitting, a transmitter $T_i$ compares the power it
senses, $P^{CS}(T_i)$, with the power threshold $P_{th}$. A key
disadvantage of this approach is that $P^{CS}(T_i)$ is a cumulative
power from all the other nodes that are concurrently transmitting.
The cumulative nature makes it impossible to tell whether
$P^{CS}(T_i)$ is from one particular nearby transmitter or a group
of far-off transmitters \cite{Kyle}. This reduces spatial reuse, as
illustrated by the example in Fig. \ref{limitspr}.

There are four links in Fig. \ref{limitspr}, with
\emph{Safe-}$CSR_{\text{\emph{cumulative}}}$ set as in
\eqref{SafeCSRangeReq}. In Fig. \ref{limitspr}, the distance
$d(T_1,T_2)$ is equal to
\emph{Safe-}$CSR_{\text{\emph{cumulative}}}$. From \eqref{eqTth}, we
f\/ind that $T_1$ and $T_2$ can not carrier sense each other, thus
they can transmit simultaneously.

First, consider the location requirement of the third link $l'_3$
that can have a concurrent transmission with both $l_1$ and $l_2$,
assuming that each transmitter can perfectly differentiate the
distances from the other transmitters. Suppose that the third link
is located on the middle line between $l_1$ and $l_2$. Based on the
carrier-sensing range analysis, the requirements are
$d(T'_3,T_1)\geq \text{\emph{Safe-}}CSR_{\text{\emph{cumulative}}}$
and $d(T'_3,T_2) \geq
\text{\emph{Safe-}}CSR_{\text{\emph{cumulative}}}$. So the third
link can be located in the position of $l'_3$, shown in Fig.
\ref{limitspr}. Furthermore, as the number of links increases, a
tight packing of the concurrent transmitters will result in a
regular equilateral triangle packing with side length
\emph{Safe-}$CSR_{\text{\emph{cumulative}}}$. The ``consumed area''
of each transmitter is a constant given by
$A=\frac{\sqrt{3}}{2}\emph{Safe-}CSR_{\text{\emph{cumulative}}}^2$.

Now, let us consider the location requirement of the third link
$l_3$ under the carrier-sensing mechanism of the current 802.11
protocol. In order to have concurrent transmissions with both $l_1$
and $l_2$, the cumulative power sensed by $T_3$ due to transmissions
of both links $l_1$ and $l_2$ should be no larger than $P_{th}$,
i.e.,
\begin{align}
P^{CS}(T_3)&=P_t\cdot d(T_3,T_1)^{-\alpha}+P_t\cdot
d(T_3,T_2)^{-\alpha}\nonumber\\
&=2\cdot P_t d(T_3,T_1)^{-\alpha}\leq P_{th},\nonumber
\end{align}
where $P_{th}$ is given in equation \eqref{abspower}. So the minimum
distance requirement on $d(T_3,T_1)$ and $d(T_3,T_2)$ is
\begin{equation}
d(T_3,T_1)=d(T_3,T_2)\geq
\left(2\frac{P_t}{P_{th}}\right)^{\frac{1}{\alpha}}=
2^{\frac{1}{\alpha}}\cdot
\text{\emph{Safe-}}CSR_{\text{\emph{cumulative}}},\nonumber
\end{equation}
as shown in Fig. \ref{limitspr}. Since $2^{\frac{1}{\alpha}}$ is
always greater than $1$, the requirement of the separation between
transmitters is increased from
$\text{\emph{Safe-}}CSR_{\text{\emph{cumulative}}}$ (i.e.,
$d(T_1,T_2)$) to
$2^{\frac{1}{\alpha}}\text{\emph{Safe-}}CSR_{\text{\emph{cumulative}}}$
(i.e., $d(T_1,T_3)$ and $d(T_2,T_3)$). The requirement on the
separation between transmitters will increase progressively as the
number of concurrent links increases, and the corresponding packing
of transmitters will be more and more sparse. As a result, spatial
reuse is reduced as the number of links increases.

Another thing to notice is that the order of the transmissions of
links also affects spatial reuse in the conventional carrier-sensing
mechanism. Consider the three links, $l_1$, $l_2$ and $l_3$ in Fig.
\ref{limitspr} again. If the sequence of transmissions is $\{l_1,
l_2,l_3\}$, as discussed above, $T_1$, $T_2$ and $T_3$ sense a power
no greater than $P_{th}$, and thus $l_1$, $l_2$ and $l_3$ can be
active simultaneously. If the sequence of transmissions on these
links is $\{l_2,l_3,l_1\}$, however, both $T_2$ and $T_3$ sense a
power no larger than $P_{th}$. But the cumulative power sensed by
$T_1$ in this case is
\begin{align}
&P^{CS}(T_1)=P_t\cdot
d(T_3,T_1)^{-\alpha}+P_t\cdot d(T_2,T_1)^{-\alpha}\nonumber\\
=&P_t
\left(2^{\frac{1}{\alpha}}\text{\emph{Safe-}}CSR_{\text{\emph{cumulative}}}\right)^{-\alpha}+P_t
\left(\text{\emph{Safe-}}CSR_{\text{\emph{cumulative}}}\right)^{-\alpha}\nonumber\\
=&\frac{3}{2}P_{th}> P_{th}.\nonumber
\end{align}
Therefore, $T_1$ will sense the channel busy and will not initiate
the transmission on $l_1$. The spatial reuse is unnecessarily
reduced because there would have been no collisions had $T_1$ decide
to transmit\footnote[8]{This corresponds to the exposed-node
phenomenon.}.


\subsection{Incremental-Power Carrier-Sensing (IPCS) Mechanism}
We propose an enhanced physical carrier-sensing mechanism called
Incremental-Power Carrier-Sensing (IPCS) to solve the issues
identif\/ied in section \ref{limitation}. Specif\/ically, the IPCS
mechanism can implement the safe carrier-sensing range accurately by
separating the detected powers from multiple concurrent
transmitters.

There are two fundamental causes for collisions in a CSMA network.
Besides hidden nodes, collisions can also happen when the backoff
mechanisms of two transmitters count down to zero simultaneously,
causing them to transmit together. Note that for the latter, each of
the two transmitters is not aware that the other transmitter will
begin transmission at the same time. Based on the power that it
detects, it could perfectly be safe for it to transmit together with
the existing active transmitters, only if the other transmitter did
not decide to join in at the same time. There is no way that the
carrier-sensing mechanism can prevent this kind of collisions. This
paper addresses the hidden-node phenomenon only. To isolate the
second kind of collisions, we will assume in the following
discussion of IPCS that no two transmitters will transmit
simultaneous\footnote[3]{Collisions due to simultaneous
countdown-to-zero can be tackled by an exponential backoff mechanism
in which the transmission probability of each node is adjusted in a
dynamic way based on the busyness of the network. In WiFi, for
example, the countdown window is doubled after each collision. The
probability of this kind of collisions can be made small with a
proper design of the backoff mechanism}. Conceptually, we could
imagine the random variable associated with backoff countdown to be
continuous rather than discrete, which means that the
starting/ending of one link's transmission will coincide with the
starting/ending of another link's transmission with zero
probability.



The key idea of IPCS is to utilize the whole carrier-sensing power
history, not just the carrier-sensing power at one particular time.
In CSMA networks, each transmitter $T_i$ carrier senses the channel
except during the time when it transmits DATA or receives ACK. The
power being sensed increases if a link starts to transmit, and
decreases if a link f\/inishes transmission. As a result, the power
sensed by transmitter $T_i$, denoted by $P_{i}^{CS}(t)$, is a
continuous function of time $t$.

In IPCS, instead of checking the absolute power sensed at time $t$,
the transmitter checks increments of power in the past up to time
$t$. If the packet duration $t_{packet}$ (including both DATA and
ACK frames and the SIFS in between) is a constant for all links,
then it suff\/ices to check the power increments during the time
window $[t-t_{packet},t]$\footnote[4]{This assumption is used to
simplify explanation only. In general, we could check a time window
suff\/iciently large to cover the maximum packet size among all
links.}. Let $\{t_1, t_2, \cdots, t_k, \cdots\}$ denote the time
instances when the power being sensed changes, and $\{\Delta
P_{i}^{CS}(t_1), \Delta P_{i}^{CS}(t_2), \cdots, \Delta
P_{i}^{CS}(t_k), \cdots\}$ denote the corresponding increments,
respectively. In IPCS, transmitter $T_i$ will decide the channel to
be \emph{idle} at time $t$ if the following conditions are met:
\begin{equation}
\Delta P_{i}^{CS}(t_k)\leq P_{th}, \quad \forall t_k \text{ such
that } t-t_{packet}\leq t_k\leq t,\label{IPCScondition}
\end{equation}
where $P_{th}$ is the carrier-sensing power threshold determined
according to $CSR$; otherwise, the channel is deemed to be
\emph{busy}. Since $\Delta P_{i}^{CS}(t_k)$ is negative if a link
stops transmission at some time $t_k$, we only need to check the
instances where the power increments are positive.



By checking every increment in the detected power, $T_i$ can
separate the powers from all concurrent transmitters, and can map
the power prof\/ile to the required distance information. In this
way, IPCS can ensure the separations between all transmitters are
tight in accordance with Theorem \ref{SafeCSRange}.


\begin{theorem}\label{theorem2}
If the carrier-sensing power threshold $P_{th}$ in the IPCS
mechanism is set as:
\begin{equation}
\label{safeIPth}
P_{th}=P_t\left(\text{\emph{Safe-}}CSR_{\text{\emph{cumulative}}}\right)^{-\alpha},
\end{equation}
where $\text{\emph{Safe-}}CSR_{\text{\emph{cumulative}}}$ is the
safe carrier-sensing range in \eqref{SafeCSRangeReq}, then it is
suff\/icient to prevent hidden-node collisions under the cumulative
interference model.
\end{theorem}

\begin{proof}
The proof is given in Appendix \ref{theorem2proof}.
\end{proof}

\begin{figure}[http]
\begin{center}
\includegraphics [height=2.7cm]{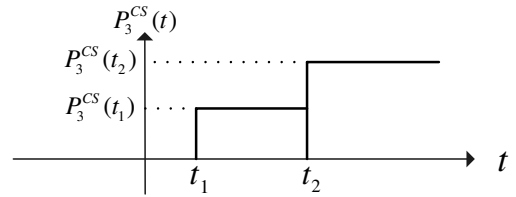}
\end{center}
\begin{center}
\vspace*{-0.20cm} \caption{The power sensed by transmitter $T'_3$ as
a function of time} \label{cspowert}
\end{center}
\end{figure}

Let us use Fig. \ref{limitspr} again to show how IPCS can implement
the safe carrier-sensing range successfully. We set the
carrier-sensing power threshold $P_{th}$ as in \eqref{safeIPth}. We
will show that the location requirement of the third link under IPCS
is the same as indicated by the safe carrier-sensing range (location
$l'_3$ in Fig. \ref{limitspr}). The transmitter of the third link
will only initiate its transmission when it senses the channel to be
idle. Its carrier-sensed power is shown in Fig. \ref{cspowert}.
Without loss of generality, suppose that link $l_1$ starts
transmission before $l_2$. The third transmitter detects two
increments in its carrier-sensed power at time instances $t_1$ and
$t_2$ which are due to the transmissions of $T_1$ and $T_2$,
respectively. In the IPCS mechanism, the third transmitter will
believe that the channel is idle (i.e., it can start a new
transmission) if the following is true:
\begin{equation}
\label{ipcscondexample}
\begin{cases}
\Delta P_{3}^{CS} (t_1 )=P_td(T'_3,T_1)^{-\alpha} \le P_{th} , \\
\Delta P_{3}^{CS} (t_2 )=P_td(T'_3,T_2)^{-\alpha} \le P_{th}.
\end{cases}
\end{equation}
Substituting $P_{th}$ in \eqref{safeIPth} to
\eqref{ipcscondexample}, we f\/ind that the requirements in
\eqref{ipcscondexample} are equivalent to the following distance
requirements:
\begin{equation}
\begin{cases}
d(T'_3,T_1)\geq\text{\emph{Safe-}}CSR_{\text{\emph{cumulative}}},\nonumber
\\
d(T'_3,T_2)\geq\text{\emph{Safe-}}CSR_{\text{\emph{cumulative}}}.\nonumber
\end{cases}
\end{equation}
So the third link can be located at the position of $l'_3$, as shown
in Fig. \ref{limitspr}, instead of far away at the location of $l_3$
as in the conventional carrier-sensing mechanism.

Compared with the conventional carrier-sensing mechanism, the
advantages of IPCS are
\begin{enumerate}
\item{IPCS is a pairwise carrier-sensing mechanism.
In the IPCS mechanism, the power from each and every concurrent link
is checked individually. This is equivalent to checking the
separation between every pair of concurrent transmission links. With
IPCS, all the analyses based on the concept of a carrier-sensing
range remain valid.}
\item{IPCS improves spatial reuse and network
throughput. In the conventional carrier-sensing mechanism, the link
separation requirement increases as the number of concurrent links
increases. In IPCS, however, the link separation requirement remains
the same. Furthermore, because IPCS is a pairwise mechanism, the
order of the transmissions of links will not affect the spatial
reuse.}
\end{enumerate}

\section{Simulations Results}\label{simulation}


We perform simulations to evaluate the relative performance of IPCS
and conventional Carrier Sensing (CS). In our simulations, the nodes
are located within in a square area of $300m \times 300m$. The
locations of the transmitters are generated according to a Poisson
point process. The length of a link is uniformly distributed between
$10$ and $20$ meters. More specif\/ically, the receiver associated
with a transmitter is randomly located between the two concentric
circles of radii $10m$ and $20m$ centered on the transmitter. We
study the system performance under different link densities by
varying the number of links in the square from $1$ to $200$ in our
simulations.

%

The simulations are carried out based on the 802.11b protocol. The
common physical layer link rate is $11Mbps$. The packet size is
$1460$ Bytes. The minimum and maximum backoff window $CW_{\min}$ and
$CW_{\max}$ are 31 and 1023, respectively. The slot time is $20\mu
s$. The SIFS and DIFS are $10\mu s$ and $50\mu s $, respectively.
The transmit power $P_t$ is set as $100mW$. The path-loss exponent
$\alpha$ is $4$, the SIR requirement $\gamma_0$ is $20$, and the
corresponding \emph{Safe-}$CSR_\text{\emph{cumulative}}$ equals
$117.6m$ based on \eqref{SafeCSRangeReq}. That is, the
carrier-sensing power threshold
$P_{th}=P_t\left(\text{\emph{Safe-}}CSR_\text{\emph{cumulative}}\right)^{-\alpha}
=5.23\times10^{-7}mW$.

%

\begin{figure}[t]
\begin{center}
\includegraphics [height=6.8cm]{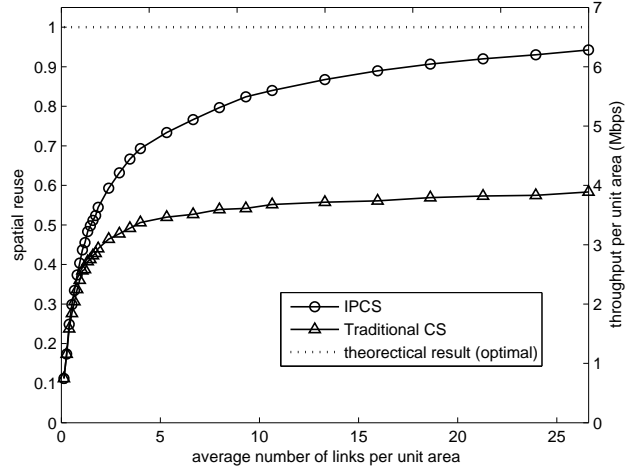}
\end{center}
\begin{center}
\vspace*{-0.20cm} \caption{Spatial reuse and network throughput
under IPCS and the conventional CS mechanisms} \label{sprthr}
\vspace*{0.40cm}
\end{center}
\end{figure}

In Fig. \ref{sprthr}, we plot spatial reuse and network throughput
under IPCS and the conventional CS mechanisms. Simulation results
show that network throughput is proportional to spatial reuse. So we
plot these two results in the same f\/igure. We def\/ine a ``unit
area'' as the ``consumed area'' of each ``active'' transmitter under
the tightest packing. Given
\emph{Safe-}$CSR_{\text{\emph{cumulative}}}=117.6m$, according to
the carrier-sensing range analysis, the ``unit area'' is
$\frac{\sqrt{3}}{2}$\emph{Safe-}$CSR_{\text{\emph{cumulative}}}^2=1.197\times10^4m^2$.
The x-axis is the average number of links (i.e., all links,
including active and inactive links) per unit area as we vary the
total number of links in the whole square. That is, the x-axis
corresponds to the link density of the network. The left y-axis is
the spatial reuse, or the average ``active'' link density in the
network. The optimal value of the spatial reuse is $1$, which is
shown as a dashed line in Fig. \ref{sprthr}. The right y-axis is the
throughput per unit area.

It is clear from Fig. \ref{sprthr} that IPCS outperforms the
conventional CS. The improvement becomes more signif\/icant when the
network becomes denser. At the densest point in the f\/igure,
spatial reuses under IPCS and conventional CS are $0.9424$ and
$0.5834$, respectively. The network throughputs per unit area are
$6.66Mbps$ and $4.08Mbps$, respectively. Using conventional CS as
the base line, the IPCS improves spatial reuse and network
throughput by more than $60\%$.

Under the conventional CS, in order to make sure the cumulative
detected power is no larger than the power threshold $P_{th}$, the
packing of concurrent transmission links will become more and more
sparse as additional number of links attempt to transmit. Under
IPCS, this does not occur. As a result, the improvement in spatial
reuse is more signif\/icant as the network becomes denser.

We also f\/ind that when the network becomes denser and denser,
spatial reuse under IPCS becomes very close to the theoretical
result. The small gap is likely due to the fact that a link which
could be active concurrently under IPCS does not exist in the given
topology. The probability of this happening decreases as the network
becomes denser.

\section{Conclusion}\label{conclusion}

In this paper, we derive a threshold on the safe carrier-sensing
range that is suff\/icient to prevent hidden-node collisions under
the cumulative interference model. We show that the safe
carrier-sensing range required under the cumulative interference
model is larger than that required under the pairwise interference
model by a constant multiplicative factor.

We propose a novel carrier-sensing mechanism called
Incremental-Power Carrier-Sensing (IPCS) that can realize the safe
carrier-sensing range concept in a simple way. The IPCS checks every
increment in the detected power so that it can separate the detected
power of every concurrent transmitter, and then maps the power
prof\/ile to the required distance information. Our simulation
results show that IPCS can boost spatial reuse and network
throughput by more than $60\%$ relative to the conventional
carrier-sensing mechanism in the current 802.11 protocol.

One future research direction is to further tighten the safe
carrier-sensing range according to the topology information. In this
paper, we have assumed a common safe carrier-sensing range for all
transmitters. Allowing the carrier-sensing range to vary from
transmitter to transmitter according to the local network
topological structures may improve spatial reuse further. In this
paper, we have not considered virtual carrier sensing (i.e., the
RTS/CTS mode in 802.11). Ensuring hidden-node free operation under
virtual carrier sensing is rather complicated even under the
pairwise interference model (see \cite{libinhdfvcs} for details.)
The study of interference-safe transmissions for virtual carrier
sensing under the cumulative interference model is a subject for
further study.

\appendices

\section{The Need for RS(Re-Start) Mode}\label{apprestartmode}

It is shown in \cite{LiBin} that although the carrier-sensing range
is suff\/iciently large for the SINR requirements of all nodes,
transmission failures can still occur due to the ``Receiver-Capture
effect''.

\begin{figure}[http]
\begin{center}
\includegraphics [height=1.4cm]{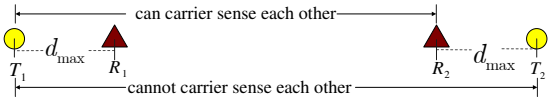}
\end{center}
\begin{center}
\caption{Collision due to ``Receiver-Capture effect''}
\label{restart}
\end{center}
\end{figure}

Take a two-link case shown in Fig. \ref{restart} as an example. In
Fig. \ref{restart}, $d(T_1,T_2)>CSR$ and $d(T_1, R_2)<CSR$. So the
transmitters $T_1$ and $T_2$ can not carrier-sense each other, but
$R_2$ can sense the signal transmitted from $T_1$. Suppose that
$CSR$ is set large enough to guarantee the SINR requirements on
$l_1$ and $l_2$ (both the DATA frames and the ACK frames). If $T_1$
transmits f\/irst, then $R_2$ will have sensed the signal of $T_1$
and the default operation in most 802.11 products is that $R_2$ will
not attempt to receive the later signal from $T_2$, even if the
signal from $T_2$ is stronger. This will cause the transmission on
link $l_2$ to fail. It is further shown in \cite{LiBin} that no
matter how large the carrier-sensing range is, we can always come up
with an example that gives rise to transmission failures, if the
``Receiver-Capture effect'' is not dealt with properly. This kind of
collisions can be solved with a receiver ``RS (Re-Start) mode''.
With RS mode, a receiver will switch to receive the stronger packet
as long as the SINR threshold ${\gamma}_0$ for the later link can be
satisf\/ied.

\section{Proof of Theorem \ref{SafeCSRange}}\label{theorem1proof}

\begin{proof}
With the receiver's RS mode, in order to prevent hidden-node
collisions in 802.11 networks, we only need to show that condition
\eqref{SafeCSRangeReq} is suff\/icient to guarantee the satisfaction
of both the SIR requirements \eqref{SINR1} and \eqref{SINR2} of all
the concurrent transmission links.

Let $\mathcal{S}^{ CS}$ denote a subset of links that are allowed to
transmit concurrently under the
\emph{Safe-}$CSR_\text{\emph{cumulative}}$ setting. Consider any two
links $l_i$ and $l_j$ in $\mathcal{S}^{ CS}$, we have
\begin{equation}
d(T_j,T_i)\geq\text{\emph{Safe-}}CSR_\text{\emph{cumulative}}=(K+2)d_{\max}.\nonumber
\end{equation}
Because both the lengths of links $l_i$ and $l_j$ satisfy
\begin{equation}
d(T_i,R_i)\leq d_{\max}, \quad d(T_j,R_j)\leq d_{\max},\nonumber
\end{equation}
we have the following based on the triangular inequality
\begin{align}
&d(T_j,R_i)\geq d(T_j,T_i)-d(T_i,R_i)\geq (K+1)d_{\max },\nonumber\\
&d(R_j,T_i)\geq d(T_i,T_j)-d(T_j,R_j)\geq (K+1)d_{\max },\nonumber\\
&d(R_j,R_i)\geq d(R_i,T_j)-d(T_j,R_j)\geq Kd_{\max }.\nonumber
\end{align}

We take the most conservative distance $Kd_{\max }$ in our
interference analysis (i.e., we will pack the interference links in
a tightest manner given the
\emph{Safe-}$CSR_\text{\emph{cumulative}}$ in
\eqref{SafeCSRangeReq}). Consider any two links $l_i$ and $l_j$ in
$\mathcal{S}^{ CS}$. The following four inequalities are
satisf\/ied:
\begin{align}
d(T_i,T_j)&\geq Kd_{\max }, \quad d(T_i,R_j)\geq Kd_{\max },\nonumber \\
d(T_j,R_i)&\geq Kd_{\max }, \quad d(R_i,R_j)\geq Kd_{\max
}.\nonumber
\end{align}

Consider any link $l_i$ in $\mathcal{S}^{ CS}$. We will show that
the SIR requirements for both the DATA frame and the ACK frame can
be satisf\/ied. We f\/irst consider the SIR requirement of the DATA
frame. The SIR at $R_i$ is:
\begin{align}
SIR=\frac{P_td^{ - \alpha }\left( {T_i ,R_i }
\right)}{\sum\limits_{l_j\in \mathcal{S}^{ CS},j \ne i} {P_td^{ -
\alpha }\left( {S_j ,R_i } \right)} }\nonumber
\end{align}

For the received signal power we consider the worst case that
$d(T_i,R_i)=d_{\max }$.  So we have
\begin{equation}
P_td^{ - \alpha }\left( {T_i ,R_i } \right)\geq P_t\cdot{d_{\max
}^{- \alpha }}.\label{sigP}
\end{equation}

To calculate the cumulative interference power, we consider the
worst case that all the other concurrent transmission links have the
densest packing, in which the link lengths of all the other
concurrent transmission links are equal to zero. In this case, the
links degenerate to nodes. The minimum distance between any two
links in $\mathcal{S}^{ CS}$ is $Kd_{\max }$. The densest packing of
nodes with the minimum distance requirement is the hexagon packing
(as shown in Fig. \ref{topo}).

If link $l_j$ is the f\/irst layer neighbor link of link $l_i$, we
have $d(S_j,R_i)\geq Kd_{\max }$. Thus we have
\begin{equation*}
{P_td^{ - \alpha }\left( {S_j ,R_i } \right)}  \leq
{P_t{\left(Kd_{\max }\right)}^{ - \alpha }}  =
\frac{1}{K^{\alpha}}\cdot P_t{d_{\max }^{- \alpha }},
\end{equation*}
and there are at most 6 neighbor links in the f\/irst layer.

If link $l_j$ is the second layer neighbor link of link $l_i$, we
have $d(S_j,R_i)\geq \sqrt{3} Kd_{\max }$. Thus we have
\begin{equation*}
{P_td^{ - \alpha }\left( {S_j ,R_i } \right)}  \leq
{P_t{\left(\sqrt{3} Kd_{\max }\right)}^{ - \alpha }}  =
\frac{1}{\left(\sqrt{3} K\right)^{\alpha}}  P_t{d_{\max }^{- \alpha
}},
\end{equation*}
and there are at most 12 neighbor links in the second layer.

If link $l_j$ is the $n$th layer neighbor link of link $l_i$ with
$n\geq 2$, we have $d(S_j,R_i)\geq \frac{\sqrt{3}}{2}n\cdot Kd_{\max
}$. Thus we have
\begin{equation*}
{P_td^{ - \alpha }\left( {S_j ,R_i } \right)}  \leq
{P_t{\left(\frac{\sqrt{3}}{2}n Kd_{\max }\right)}^{ - \alpha }}
=\frac{1}{\left(\frac{\sqrt{3}}{2}n K\right)^{\alpha}} P_t{d_{\max
}^{- \alpha }},
\end{equation*}
and there are at most $6n$ neighbor links in the $n$th layer.

So the cumulative interference power satisf\/ies:
\begin{align}
&{\sum\limits_{l_j\in \mathcal{S}^{ CS},j \ne i} {P_td^{ - \alpha
}\left(
{S_j ,R_i } \right)} }\nonumber\\
\leq &\left(6 \cdot \left( {\frac{1}{K}} \right)^\alpha \mbox{ +
}\sum\limits_{n = 2}^\infty {6n\left( {\frac{2}{\sqrt 3 nK}}
\right)^\alpha }\right)\cdot P_t{d_{\max }^{- \alpha }} \nonumber\\
= & 6 \cdot \left( {\frac{1}{K}} \right)^\alpha \left( {\mbox{1 +
}\sum\limits_{n = 2}^\infty {n\left( {\frac{2}{\sqrt 3 n}}
\right)^\alpha } } \right)\cdot P_t{d_{\max }^{- \alpha }}\nonumber\\
= &6 \cdot \left( {\frac{1}{K}} \right)^\alpha \left( {\mbox{1 +
}\left( {\frac{2}{\sqrt 3 }} \right)^\alpha \sum\limits_{n =
2}^\infty {n\left( {\frac{1}{n}} \right)^\alpha } } \right)\cdot
P_t{d_{\max }^{- \alpha }}\nonumber\\
= &6 \cdot \left( {\frac{1}{K}} \right)^\alpha \left( {\mbox{1 +
}\left( {\frac{2}{\sqrt 3 }} \right)^\alpha \sum\limits_{n =
2}^\infty {\frac{1}{n^{\alpha - 1}}} } \right)\cdot P_t{d_{\max }^{-
\alpha }}\nonumber\\
\le & 6 \cdot \left( {\frac{1}{K}} \right)^\alpha \left( {\mbox{1 +
}\left( {\frac{2}{\sqrt 3 }} \right)^\alpha \frac{1}{\alpha - 2}}
\right)\cdot P_t{d_{\max }^{- \alpha }}\label{boundrie}\\
=&\frac{P_t{d_{\max }^{- \alpha }}}{\gamma_0},\label{follK}
\end{align}
where \eqref{boundrie} follows from a bound on Riemann's zeta
function and \eqref{follK} follows from the def\/inition of $K$ in
\eqref{KReq}.

\begin{figure}[http]
\begin{center}
\vspace*{0.4cm}
\includegraphics [height=5.5cm]{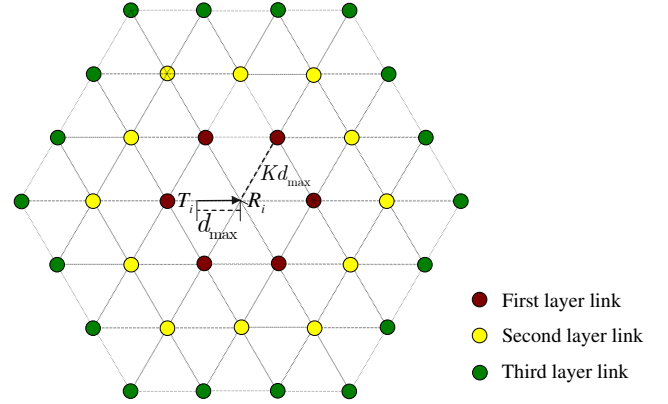}
\end{center}
\begin{center}
\caption{The packing of the interfering links in the worst case}
\label{topo}
\end{center}
\end{figure}

According to \eqref{sigP} and \eqref {follK}, we f\/ind that the SIR
of the DATA frame of link $l_i$ at the receiver $R_i$ satisf\/ies:
\begin{equation}
SIR=\frac{P_td^{ - \alpha }\left( {T_i ,R_i }
\right)}{\sum\limits_{l_j\in \mathcal{S}^{ CS},j \ne i} {P_td^{ -
\alpha }\left( {S_j ,R_i } \right)} } \geq \frac{P_t\cdot{d_{\max
}^{- \alpha }}}{\frac{P_t{d_{\max }^{- \alpha
}}}{\gamma_0}}=\gamma_0.\nonumber
\end{equation}

This means that the SIR requirement of the successful transmission
of the DATA frame on link $l_i$ can be satisf\/ied.

The proof that the SIR requirement of the ACK frame on link $l_i$
can be satisf\/ied follows a similar procedure as above. So for any
link $l_i$ in the concurrent transmission link set $\mathcal{S}^{
CS}$, condition \eqref{SafeCSRangeReq} is suff\/icient to satisfy
the SIR requirements of the successful transmissions of both its
DATA and ACK frames. This means that, together with the receiver's
RS mode, condition \eqref {SafeCSRangeReq} is suff\/icient for
preventing hidden-node collisions in CSMA networks under the
cumulative interference model.
\end{proof}

\section{Proof of Theorem \ref{theorem2}}\label{theorem2proof}
\begin{proof}
Consider any link $l_i$ in the link set $\mathcal{L}$. Transmitter
$T_i$ will always do carrier sensing except when it transmits DATA
frame or receives ACK frame. We show that condition \eqref{safeIPth}
is suff\/icient to prevent hidden-node collisions in the following
two situations, which cover all the possible transmission scenarios:
\begin{enumerate}
\item{Link $l_i$ has monitored the channel for at least $t_{packet}$
before its backoff counter reaches zero and it
transmits.}\label{proofcase1}
\item{Link $l_i$ f\/inishes a transmission; then monitors the channel
for less than $t_{packet}$ when its backoff counter reaches zero;
then it transmits its next packet.}\label{proofcase2}
\end{enumerate}

Let us f\/irst consider case \ref{proofcase1}$)$:

We show that for the links that are allowed to transmit
simultaneously, the separation between any pair of transmitters is
no less than the safe carrier-sensing range
$\text{\emph{Safe-}}CSR_\text{\emph{cumulative}}$. We use inductive
proof method. Suppose that before $l_i$ starts to transmit, there
are already $M$ links transmitting and they are collectively denoted
by the link set $\mathcal{S}^{ CS}$. Without loss of generality,
suppose that these $M$ links begin to transmit one by one, according
to the order $l_1,l_2,\cdots,l_M$. For any link $l_j\in\mathcal{S}^{
CS}$, let $t_j$ and $t'_j$ denote the times when link $l_j$ starts
to transmit the DATA frame and the ACK frame, respectively.

In our inductive proof, by assumption we have 
\begin{equation}
d(T_j,T_k)\geq \text{\emph{Safe-}}CSR_\text{\emph{cumulative}},
\forall j,k\in\{1,\cdots,M\}, j\neq k.\label{trandis}
\end{equation}
We now show that condition \eqref{trandis} will still hold after
link $l_i$ starts its transmission.

Before link $l_i$ starts its transmission, transmitter $T_i$
monitors the channel for a time period of $t_{packet}$. So $T_i$ at
least senses $M$ increments in the carrier-sensing power
$P_{i}^{CS}(t)$ that happen at time $t_1,t_2,\cdots,t_M$ when the
links in $\mathcal{S}^{ CS}$ start to transmit their DATA frames.
There may also be some increments in the $P_{i}^{CS}(t)$ that happen
at $t'_1,t'_2,\cdots,t'_M$ if the links in $\mathcal{S}^{ CS}$ start
to transmit the ACK frames before link $l_i$ starting it
transmission. In the IPCS mechanism, at least the following $M$
inequalities must be satisf\/ied if $T_i$ can start its
transmission:
\begin{equation}
\Delta P_{i}^{CS} (t_j ) \le P_{th}, \quad \text{for} \quad
j=1,\cdots,M. \nonumber
\end{equation}
Because
\begin{align}
&\Delta P_{i}^{CS} (t_j )=P_td(T_i,T_j)^{-\alpha}, \nonumber\\
&P_{th}=P_t\left(\text{\emph{Safe-}}CSR_\text{\emph{cumulative}}\right)^{-\alpha},\nonumber
\end{align}
we have
\begin{equation}
d(T_i,T_j)\geq \text{\emph{Safe-}}CSR_\text{\emph{cumulative}} \quad
\text{for} \quad j=1,\cdots,M. \nonumber
\end{equation}
Thus, we have shown that the separation between any pair of
transmitters in the link set $\mathcal{S}^{ CS}\cup l_i$ is no less
than $\text{\emph{Safe-}}CSR_\text{\emph{cumulative}}$ after link
$l_i$ starting transmission.


Now let us consider case \ref{proofcase2}$)$:

Before starting the transmission of the $(m+1)$th packet, link $l_i$
f\/irst f\/inishes the transmission of the $m$th packet (from time
$t_i(m)$ to $t_i(m)+t_{packet}$), and waits for a DIFS plus a
backoff time (from time $t_i(m)+t_{packet}$ to $t_i(m+1)$). Let
$\mathcal{S}^{ CS}$ denote the set of links that are transmitting
when $l_i$ starts the $(m+1)$th packet at time $t_i(m+1)$. Consider
any link $l_j$ in set $\mathcal{S}^{ CS}$. Because the transmission
time of every packet in the network is $t_{packet}$. We know that
the start time $t_j$ of the concurrent transmission on link $l_j$
must range from $t_i(m)$ to $t_i(m+1)$, i.e., $t_i(m)<t_j<t_i(m+1)$.

If $t_i(m)+t_{packet}< t_j<t_i(m+1)$, this means $t_j$ is in the
DIFS or the backoff time of link $l_i$. During this period,
transmitter $T_i$ will do carrier sensing. The IPCS mechanism will
make sure that the distance between $T_i$ and $T_j$ satisf\/ies
$d(T_i,T_j)\geq \text{\emph{Safe-}}CSR_\text{\emph{cumulative}}$.

If $t_i(m)< t_j<t_i(m)+t_{packet}$, this means $t_j$ falls into the
transmission time of the $m$th packet of link $l_i$. During the
transmission time, $T_i$ is not able to do carrier sensing because
it is in the process of transmitting the DATA frame or receiving the
ACK frame. However, the transmitter $T_j$ will do carrier sensing
before it starts to transmit at time $t_j$. The carrier sensing done
by $T_j$ can make sure that the distance between $T_i$ and $T_j$
satisf\/ies $d(T_i,T_j)\geq
\text{\emph{Safe-}}CSR_\text{\emph{cumulative}}$.

So for any link $l_j$ in $\mathcal{S}^{ CS}$, we have
$d(T_i,T_j)\geq \text{\emph{Safe-}}CSR_\text{\emph{cumulative}}$.


\end{proof}

\end{document}